# Restricted Repetitive Behaviors in Adolescent Males with Autism: Volatility in Brain Functional Connectivities

Gerardo Noriega [1]

*Abstract*— This paper studies subtypes of restricted, repetitive and stereotypical behaviors (RRBs) in adolescent males with autism spectrum disorder (ASD) from the viewpoint of the dynamics of brain functional connectivities (FCs). Data from the ABIDE-II repository and Repetitive Behavior Scale-Revised (RBS-R) metrics are used to form two ASD groups with tightly controlled demographics; one comprises subjects with scores above threshold for the self-injurious behaviors (SIBs) subscale, and the other subjects with scores below threshold for SIBs, but above threshold for at least one of the other subscales (stereotyped, compulsive, ritualistic, insistence on sameness, restricted interests). The dynamics of the coherence for FCs across distinct frequency bands are compared against matched controls, using a novel volatility measure computed in time-frequency space. We find statistically significant differences, on average, in the volatility of a relatively small set of FCs, most mapping to either the default mode network or the cerebellum, in the mid- and high-frequency bands, and yielding higher volatility in subjects with high levels of SIBs. Results suggest a distinct underlying profile for SIBs involving multiple brain regions associated with rewards and emotions processing. The work contributes to the identification of neural substrates potentially underlying behavioral subtypes, and may help target interventions.

## I. Introduction

Restrictive, repetitive and stereotypical patterns of behavior (RRBs) have always played an important role in autism spectrum disorder (ASD), but only recently have been given some emphasis in the literature. Sensory issues, for example, inherently intertwined with RRBs and identified as relevant going back to Kanner's original work [1], were not given formal recognition in the diagnosis criteria of ASD until the release of the Fifth Edition of the Diagnostic and Statistical Manual of Mental Disorders [2].

The term 'RRBs' broadly refers to behaviors which involve repetition, invariance, and lack of flexibility and an obvious purpose, all within the framework of highly-restricted and fixated interests. They have broad negative impact on an individual's ability to communicate, engage in other activities, and establish social relationships. In severe cases they lead to anxiety, aggression and self-injurious behaviors. Classification criteria remain controversial, with rare consensus on clinicians' evaluations [3]. Arguably, two of the best assessment tools available are the Autism Diagnostic Interview-Revised (ADI-R) [4], and the Repetitive Behavior Scale-Revised (RBS-R) [5]. The former has been well-established for quite some time, but the factors it includes in the RRB domain are scarce and concentrated in restricted interests and behaviors. On the other hand, the latter assesses a wide variety of RRBs with similarly well-accepted validity and reliability.

RBS-R is an informant-based scale comprising 43 items conceptually grouped into six subscales: (i) stereotyped, (ii) self-injurious, (iii) compulsive (i.e., repeated and performed according to a rule), (iv) ritualistic (daily activities done in a similar manner), (v) insistence on sameness (resistance to change), and (vi) restricted interests. Items are scored on a four-point Likert scale: (0) behavior does not occur; (1) occurs and is a mild problem; (2) occurs and is a moderate problem; (3) occurs and is a severe problem. The scale captures the breadth of RRBs with much finer granularity than other alternatives, including the Restricted, Repetitive, and Stereotyped Patterns of Behavior Subscore of the ADI-R tool (ADI-R-RRB) [4]. Unfortunately, RBS-R did not become widely available in large, multi-site, public data sets until recently. It is included, for example, for a substantial (~44%) subset of the subjects with autism diagnosis in the ABIDE-II [6] repository, but not in its predecessor ABIDE-I.

Copious research in a broad range of fields (including genetics, psychology, neuroanatomy, neurophysiology, computational modeling, and neuroimaging) has yielded important advances in the understanding of ASD. In particular, functional connectivity (FC) magnetic resonance imaging (fMRI) studies have played a pivotal role in establishing associations between ASD and atypical brain networks. Overall findings on FC in ASD have been mixed. Per the recent review in [7] approximately half of the studies identified both over-connectivities and under-connectivities in ASD, roughly one-third exclusively under-connectivities, and the rest only over-connectivities. In contrast, an earlier review [8] found most of the studies (~75%) identified only under-connectivities, with only a couple yielding exclusively over-connectivities. The dynamic (time-varying) behavior of FCs in fMRI, in particular, has received increasing attention in the past decade [9], with substantial focus on ASD [10-16].

While research emphasis on RRB aspects of ASD has been somewhat limited, three reviews provide good context: [17] is slightly outdated but still quite comprehensive, while the more recent [18] and [3] emphasize, respectively, underlying etiology and developmental trajectory, and therapeutic approaches. The relationship between scores in the ADI-R and RBS-R measures was examined in [19], finding support for repetitive sensory motor, and insistence on sameness subcategories. Recent work on the subject includes [20], based on ABIDE-II and RBS-R metrics, with emphasis on age-related changes in RRBs, [21] which models the balance between RRBs and social communications with highly-accurate subtype predictions, and [22] which studies RRBs and FCs in infants at risk for developing ASD.

[1] https://orcid.org/0000-0003-0921-8527
G. Noriega is with RMS Instruments, Mississauga, Ontario, Canada (g.noriega@ieee.org).

A very limited subset of all research on ASD in general, and RRBs in particular, includes subjects in the "severe" end of the spectrum [23]. This may be, arguably, an unfortunate consequence of the broadening of the definition of ASD [24]. This bias in research toward the mild cases, coupled with the inherent difficulties in conducting tests such as fMRI and EEG procedures on subjects with severe sensory and/or behavioral issues, leads to alarming consequences as those most in need are grossly underrepresented in studies.

In this work we use dynamic FC analysis to better understand differences in subtypes of RRBs, identifying neural correlates of behavioral subtypes which could potentially help in developing targeted interventions. We propose a novel measure of the volatility of FCs in the time-frequency domain, based on an earlier version introduced in [25]. In a modest effort to alleviate the under-representation of individuals on the severe-end of the spectrum we emphasize self-injurious behaviors. The analysis is based on data from the ABIDE-II repository and RBS-R metrics using the original six subscales of [5], which for brevity we refer to as SS1–SS6. The work expands on preceding efforts [25-26], both based on the ABIDE-I data set and ADI-R-RRB metrics. In the former static correlations and coherence, as well as a windowed-correlations approach, were used to identify significant differences in FC strength. The dynamic behavior of the relevant FCs thus identified was further analyzed in the latter.

II. METHODS

A. General Design and Participants

Out of 521 subjects with autism diagnosis in the ABIDE-II data set, 229 have full RBS-R scores. Potential participants from this subset were restricted to right-handed males, 10–19 years of age, and under no medication. A further restriction eliminated sites with limited representation (< 4 subjects), yielding a total of 47 subjects from 4 sites: Kennedy Krieger Institute ($N_{KKI}$=13), San Diego State University ($N_{SDSU}$=6), Trinity Centre for Health Sciences ($N_{TCD}$=18), University of California, Davis ($N_{UCD}$=10).

When considering behavior severity for subscales SS1–SS6, the four-point Likert scale used [5] implies "boundaries" at 1/3 (mild–moderate) and 2/3 (moderate–severe) of each maximum score. For lack of an established precedent (to the best of the author's knowledge), a threshold for a *substantial* behavior was established at one-third of the former, i.e., [MaxScore/9]. This excludes cases where the behavior is either not present or very mild. (Noteworthy, the distributions of scores for all six subscales are heavily biased toward the low-end of the range. A threshold any higher than the one employed would yield very limited sample sizes.)

Initial analyses showed significant overlap between subsets of ASD subjects with substantial behaviors for subscales SS1, SS3, SS4, SS5 and SS6 (70–93%), while the subset for SS2 was clearly distinct from all others (33–56% overlap). This motivated stratification into two groups: (a) group 'SS2', comprising all ASD subjects with score at or above threshold for RBS-R subscale SS2 ($N_{SS2}$=18), and (b) group 'SS0', including ASD subjects with score below threshold for subscale SS2, but at or above threshold for any of subscales SS1, SS3, SS4, SS5 or SS6 ($N_{SS0}$=29) – Table I.

A total of 50 controls (CTL) were selected at random but restricted to the same set of sites as ASD, and meeting the same criteria for age, sex, handedness and medication status. Controls were further restricted to a PIQ not exceeding by more than 1.5 SDs the mean PIQ of ASD subjects.

Per Table I, there are no significant differences ($p < 0.05$) for either age or PIQ, (a) when comparing the ASD group as a whole and CTL, or (b) when comparing each of subgroups SS0 and SS2, and CTL. As would be expected, there are significant differences in VIQ (excepting SS0 vs SS2), and in FIQ (excepting SS0 vs CTL, and SS0 vs SS2).

Given the generally-accepted association between age and severity of RRBs [27], we implemented linear regression models for SS0 and SS2 scores as a function of age, finding no correlation: $R^2 < 0.007$ and 0.06, respectively.

B. Data Preprocessing

Preprocessing (*CONN Functional Connectivity SPM Toolbox, 2017*) included functional realignment and unwarp with b-spline coregistration and resampling to first scan; slice-timing correction with sinc-interpolation to match middle of acquisition time; outlier identification (displacement > 0.9mm or changes > 5 SDs); segmentation and normalization (MNI space, grey matter, white matter and CSF tissue); functional smoothing (spatial convolution, Gaussian kernel, 8mm full-width half-maximum), linear regression denoising (components from white matter and cerebrospinal areas, 12 estimated subject-motion parameters, scrubbing, session & task effects), and temporal band-pass filtering (0.008 to 0.1 Hz).

Mean region-of-interest (ROI) time-series were derived from functional data registered in standard space for the Automated Anatomical Labeling (AAL) atlas [28], fractioned to 3x3x3-mm resolution using nearest-neighbor interpolation.

C. Data Processing

We study the dynamics of the coherence between ROIs in functional connectivities (FCs), across distinct frequency bands, using a variation of the *volatility* measure introduced in [25]. For reasons that shall become clear shortly we denote the proposed measure *normalized volatility*.

The approach is based on the wavelet transform coherence (WTC), which is well-suited to study the non-stationary behavior of fMRI signals in time and frequency domains [29].

The continuous wavelet transform is defined as

$$W^X(n,s) = \left(\frac{T}{s}\right)^{\frac{1}{2}} \sum_{k=1}^{N} x_n \psi_0^*[(k-n)(T-s)], \quad (1)$$

Table I – Demographics and phenotypical information.

| Diagnostic group | Autism (ASD), Control (CTL) | | ASD | CTL |
|---|---|---|---|---|
| Gender | Male | $N_{KKI}$ | 13 | 26 |
| Age at time of scan (years) | 10–19 | $N_{SDSU}$ | 6 | 11 |
| Currently taking medications? | No | $N_{TCD}$ | 18 | 9 |
| Handedness category | Right | $N_{UCD}$ | 10 | 4 |
| | | $N_{TOT}$ | **47** | **50** |

SS2 ($N_{SS2}$ = 18): ASD subjects with RBS-R Subscale 2 at or above threshold (2/24).
SS0 ($N_{SS0}$ = 29): ASD subjects with subscale 2 below threshold, but any of RBS-R Subscales 1, 3, 4, 5 or 6 at or above threshold (2/18, 2/24, 2/18, 3/33, 1/12).

| | ASD vs CTL | SS0 vs CTL | SS2 vs CTL | SS0 vs SS2 |
|---|---|---|---|---|
| Age | ASD: 13.2 (2.5)<br>CTL: 12.6 (2.3)<br>F = 1.58<br>p = 0.2126 | SS0: 13.2 (2.7)<br>CTL: 12.6 (2.3)<br>F = 1.12<br>p = 0.2933 | SS2: 13.2 (2.3)<br>CTL: 12.6 (2.3)<br>F = 1.00<br>p = 0.3200 | SS0: 13.2 (2.7)<br>SS2: 13.2 (2.3)<br>F = 0.00<br>p = 0.9695 |
| FIQ | ASD: 105.3 (14.3)<br>CTL: 112.0 (11.9)<br>F = 6.20<br>p = 0.0145 * | SS0: 106.1 (15.2)<br>CTL: 112.0 (11.9)<br>F = 3.57<br>p = 0.0626 | SS2: 104.0 (13.1)<br>CTL: 112.0 (11.9)<br>F = 5.63<br>p = 0.0206 * | SS0: 106.1 (15.2)<br>SS2: 104.0 (13.1)<br>F = 0.24<br>p = 0.6241 |
| VIQ | ASD: 105.2 (15.9)<br>CTL: 115.3 (13.0)<br>F = 11.77<br>p = 0.0009 * | SS0: 105.4 (16.4)<br>CTL: 115.3 (13.0)<br>F = 8.80<br>p = 0.0040 * | SS2: 104.9 (15.6)<br>CTL: 115.3 (13.0)<br>F = 7.61<br>p = 0.0075 * | SS0: 105.4 (16.4)<br>SS2: 104.9 (15.6)<br>F = 0.01<br>p = 0.9196 |
| PIQ | ASD: 106.9 (14.7)<br>CTL: 107.3 (11.4)<br>F = 0.02<br>p = 0.8914 | SS0: 107.7 (14.3)<br>CTL: 107.3 (11.4)<br>F = 0.02<br>p = 0.8853 | SS2: 105.7 (15.5)<br>CTL: 107.3 (11.4)<br>F = 0.22<br>p = 0.6381 | SS0: 107.7 (14.3)<br>SS2: 105.7 (15.5)<br>F = 0.22<br>p = 0.6449 |

Entries for Age, FIQ (full-scale IQ standard score), VIQ (verbal IQ), and PIQ (performance IQ) are mean (std. dev.). * Significant at $p < 0.05$. RBS-R subscales: 1 (stereotyped), 2 (self-injurious), 3 (compulsive), 4 (ritualistic), 5 (sameness), 6 (restricted). KKI: Kennedy Krieger Inst. ($T_R$=2.5s); SDSU: San Diego State Univ. (2.0s); TCD: Trinity Centre for Health Sc. (2.0s); UCD: Univ. of California Davis (2.0s).

where $x_n$ is an $N$-length time series sampled with uniform period $T$, $n$ the time index, and $s$ the wavelet scale. The Morlet complex function $\psi_0(\eta) = \pi^{-1/4} e^{iw_0\eta} e^{-\eta^2/2}$, with $w_0 = 6$, yields a good time/frequency balance. The cross-wavelet transform for two signals $x_n$ and $y_n$, and the magnitude-squared WTC, are then given by (2) and (3), respectively:

$$W^{XY}(n,s) = W^X(n,s) W^{Y^*}(n,s) \quad (2)$$

$$R^2(n,s) = \frac{|s^{-1}W^{XY}(n,s)|^2}{|s^{-1}W^X(n,s)|^2 |s^{-1}W^Y(n,s)|^2}, \quad (3)$$

with $0 \leq R^2(n,s) \leq 1$. The magnitude-squared WTC is effectively a "correlation coefficient" in time/frequency space. The underlying algorithms [30] generate output for a frequency range dependent on the scan's repetition rate $T_R$, and number of samples $N$. The sites considered had scans with $T_R = 2$ or 2.5 s, and lengths ranging between 151 and 210 samples. Only the largest feasible range of scales common to all data sets was considered, $s_m \leq s \leq s_M$ (with $s_m = 5.17/T_R$, $s_M = 104.95/T_R$).

Calculation of the WTC was based on [30]. At discrete points in the time-frequency plane outside a conical contour in an area referred to as the *cone of influence*, $C$, computed values of the WTC have lower confidence. Per [25], let $D$ denote the set of discrete points ($n$, $s$) in the time-frequency plane which satisfy (i) $s_m \leq s \leq s_M$, and (ii) $(n,s) \notin C$. We consider frequency bands per the decomposition of fMRI time series into intrinsic mode functions (IMF) of [31]: (i) high-frequency (HF, $0.05 \leq f \leq 0.12$ Hz), with a subset of discrete points $D_{HF} \subset D$ and $card(D_{HF}) = n_{HF}$; (ii) mid-frequency (MF, $0.025 \leq f < 0.05$ Hz), with $D_{MF} \subset D$ and $card(D_{MF}) = n_{MF}$; and (iii) low-frequency (LF, $0.007 \leq f < 0.025$ Hz), with $D_{LF} \subset D$, $card(D_{LF}) = n_{LF}$ (and $D_{HF} \cup D_{MF} \cup D_{LF} = D$). A couple of points to note: (a) HF, MF and LF are equivalent to IMF2, IMF3 and IMF4 in [31], respectively; (b) to avoid discarding discrete frequencies right around the band-pass edges, the high-end of HF and the low-end of LF are slightly beyond the −3dB cut-offs used in the preprocessing stage.

The normalized volatility is based on fourth-difference signals across time, calculated at all valid scales within a frequency band. The fourth-difference at scale $s$ is a function of time given by

$$F_s(n) = [R^2(n-2,s) - 4R^2(n-1,s) + 6R^2(n,s) - 4R^2(n+1,s) + R^2(n+2,s)]/T_R^4. \quad (4)$$

Direct averaging of the magnitude of fourth-difference values at off-cone-of-influence discrete points in the time-frequency plane, per [25], yields measures fairly sensitive to frequency. This was established empirically in initial attempts to further explore the results of [25] (ABIDE-I) on a different data set (ABIDE-II, per present work). We thus propose an approach which compensates for inherent scaling effects in (4), "normalizing" volatility measures across the full frequency span and leading to more consistent results – see Section III.c. We first define *raw* volatilities at frequencies $f$ as:

$$v_f = \frac{1}{n_f} \sum_{n \in NF} |F_f(n)|, \quad (5)$$

where $NF$ is the range of 'valid' (outside-cone-of-influence) time points at frequency $f$, $n_f$ is the total number of such samples, and $F_f(n)$ is the fourth difference time series at frequency $f$ (with $f = 1/sT_R$). Raw volatilities are then normalized by a factor $f^4$, which compensates for the scaling inherent in (4), and *normalized volatilities* over frequency bands $B = \{HF, MF, LF\}$ are defined as:

$$V^B = \frac{1}{n_B}\sum_{f \in B} v_f / f^4, \quad (6)$$

where $n_B$ denotes the number of discrete frequencies within $B$. For brevity, in what follows we use simply *volatility* to refer to the measures in (6).

Fig. 1 illustrates aspects of the computational pipeline using as an example the FC between right middle frontal gyrus, orbital part, and left lobule VII of the cerebellum, with significantly higher volatility for SS2 relative to CTL in the MF band. Note how the fourth-difference (Figs. 1d, 1e) accentuates changes in coherence which appear otherwise subtle in the magnitude-squared WTC (Figs. 1b, 1c). In turn, the volatility measure, which averages across frequencies within each band (6), will yield values directly correlated with the stability of the coherence.

### D. Statistical Analysis

Notwithstanding the tightly controlled demographics of the study, two potential confounding factors were of concern: (a) scan site, as pooling data acquired with different MRI scanners may be inherently problematic, and (b) age, which despite the fairly narrow range and matching between groups remains of interest given its well-established association with the severity of RRBs, e.g., [27].

In a first stage, for each frequency band a linear regression model was built for the volatility measure, with group (SS0, SS2, CTL), scan-site (KKI, SDSU, TCD, UCD), and age as independent variables. All possible FCs amongst the 116 ROIs in the AAL atlas were considered. For initial qualification FCs had to satisfy: (a) group being a significant predictor at $p = 0.015$, (b) neither scan-site nor age being significant at $p = 0.05$, and (c) all volatility measures meeting normality criteria per Kolmogorov-Smirnov. In a second stage one-way ANOVAs were conducted followed by multiple-comparison tests (Tukey's HSD test). Potential relevant FCs were further restricted to meet: (d) significance at $p < 0.015$, (e) large effect size, |Cohen's $d$| $> 0.8$, and (f) t-test power $> 0.7$.

To ensure stability and robustness the process was repeated under a re-sampling scheme with 100 runs, 2 randomly-selected subjects excluded from each of SS0, SS2 and CTL for each run. Only FCs satisfying the conditions above for at least 40% of the runs were qualified as *relevant FCs*. The threshold at 40%, just below the inter-quartile range (IQR) of the distribution of incidence rates, was selected empirically so as to yield a good balance between ensuring that qualified FCs are stable (by avoiding connections that may appear due to random chance), while not being overly restrictive.

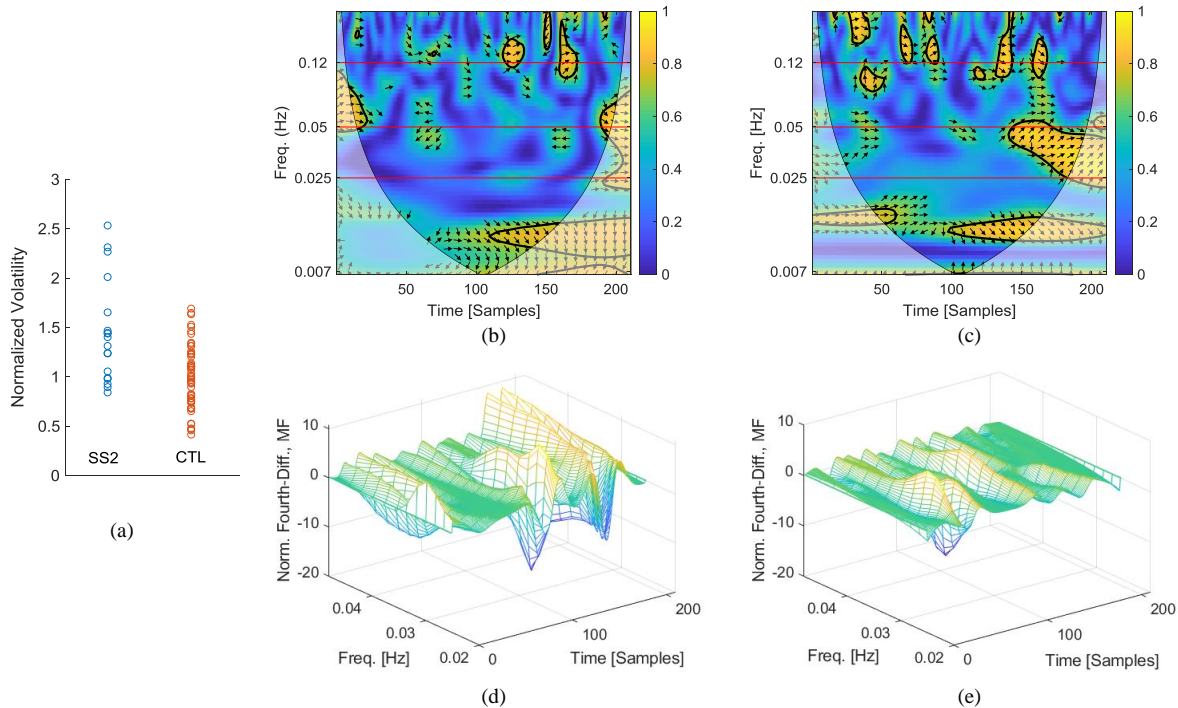

Fig. 1 – Illustration of computational pipeline for FC MFGorb.R–CBL7.L (10–101), SS2 > CTL, MF band ($p = 0.0004$, Cohen's $d = 1.13$, t-test power = 0.94). (a) Scatter plot of volatility measure for subjects in SS2 and CTL. For illustration we use one subject from SS2 with "high" volatility (2.2664), and one from CTL with "low" volatility (1.0340), both from site TCD, $T_R = 2$ s. (b) & (c) Magnitude-squared WTC for high-volatility subject (left), and low-volatility subject (right). Note the contour delimiting the cone of influence, and the markers separating frequency bands. The arrows provide a rough indication of the behavior of the WTC's phase. (d) & (e) Normalized fourth differences in the time-frequency plane, based on which the volatility measures (6) are calculated; high-volatility subject (left) and low-volatility subject (right).

## III. Results

A few notes on terminology: (a) For statistical purposes when mapping ROIs part of relevant FCs to specific brain networks, only one instance of any ROI pair with multiple occurrences is included; (b) when referring to specific FCs we use mnemonics per the AAL atlas – see [28].

### A. General

Table II summarizes all relevant FCs. For brevity, only the effect size for each connection is shown – detailed statistics are provided in Appendix A. Note that effect size (ES) and power (P) correlate well with each other, and with the incidence rate (IR) under randomized resampling: $\rho_{ES,P}=0.81$, $\rho_{ES,IR}=0.79$, $\rho_{P,IR}=0.86$. The spatial mapping of the relevant FCs is shown in Fig. 2.

The analysis yields 31 relevant FCs, with the vast majority in the MF (58.1%) and HF (38.7%) bands, and a single one in LF. For a large number of them (21/31) volatility is greater in SS2 relative to CTL (SS2 > CTL), with a subset of 3 which also yield SS2 > SS0. An additional 6 FCs yield SS2 > SS0 (but not SS2 > CTL), while 3 FCs show SS2 < SS0. A single FC yields differences between SS0 and CTL (SS0 > CTL).

The more 'severe' of any two groups compared (SS2 vs SS0 or CTL, and SS0 vs CTL) always has higher volatility, with three exceptions: SMG.R–CBL7.L (HF), TPOmed.R–CBL9.L (MF), and CBL9.L–CBLv8 (HF), with SS0 > SS2.

Two of the relevant FCs are exceptionally robust: MFGorb.R–CBL7.L and MedFGorb.R–TPOmed.L (both MF, SS2 > CTL), which prevail in 100% and 97% of the runs under randomized re-sampling, respectively – Appendix A.

### B. Mapping to Brain Networks

The vast majority of the relevant ROIs map to either the DMN or the cerebellum – Table III. For FCs in the HF band roughly half of the ROIs are in the cerebellum and one-third in DMN, whereas for MF those proportions are reversed. Interestingly, no ROIs map to the sensorimotor network, and very few to the visual or salience networks.

From a different perspective, around one-third (11/31) of the relevant FCs are between the cerebellum and parts of either the *emotions processing circuit* (comprising portions of the prefrontal cortex (SFGorb, IFGorb, OLF, SFGmed, MedFGorb, REC), the cingulate gyrus (ACG, MCG, PCG), the amygdala (AMYG) and the thalamus (THA) [27]) or the *reward circuit* (CAU, REC, AMYG, HIP, nucleus accumbens (CAU, PUT), ventral tegmental area, prefrontal cortex). Most of the FCs involve at least one of these areas: 12/12 for the HF band, and 15/18 for MF (overall, 27/31).

Longer connections are associated with FCs for which SS2 has higher volatility, increasing from an average of 80.2 mm when relative to CTL only, to 99.9 mm when relative to SS0 only, and up to 120.2 mm when relative to both CTL and SS0. For FCs with higher volatility for SS0 average distances are only 76.8 mm (SS0 > SS2), and 28.1 mm (single case with SS0 > CTL). On the other hand, there is no clear difference when comparing connection distances between relevant FCs in the HF and MF bands (83.9 and 88.3 mm, respectively).

### C. Reference to Preceding Results

Around 40% (HF) and 50% (MF) of the relevant FCs for full-size groups identified under the proposed method are common to those that the approach from [25] would yield. Most of the relevant FCs which persist under randomized re-sampling have at least one instance under equivalent re-sampling per the original scheme: 9/13 for HF, 19/20 for MF, 1/1 for LF. While the fairly high degree of agreement between the approaches is reassuring, it is important to highlight the advantages of the proposed normalized volatility:

- All FCs over threshold under randomized re-sampling are relevant for full-size groups. On the other hand, the measure from [25] yields three FCs above threshold which do not qualify as relevant for full-size groups. This particular aspect suggests that the non-normalized measure is rather sensitive, easily introducing spurious effects under minor changes to stratification.

- The normalized volatility yields more robust results: close to 90% of all relevant FCs identified for full-size groups persist under randomized re-sampling, compared to ~80%.

- It is drastically more selective in the LF band, identifying a single relevant FC, compared to 13.

Also of interest, comparison against the results from [26], which studied differences in FC *strength* between severe and mild cases of RRBs and controls, shows that close to 20% of the ROIs in relevant FCs for the present study (OLF.L, HIP.R, MTG.L, MTG.R, CBL6.R, CBLv12, CBLc1.L) were also found relevant in its predecessor. In particular, three FCs are either exact matches (OLF.L–HIP.R), or matches with one contralateral ROI (OLF.L–HIP.R vs OLF.R–HIP.R, and AMYG.R–CBLv12 vs AMYG.L–CBLv12).

## IV. Discussion and Conclusions

The majority of the relevant FCs (27/31, Table II) reveal higher volatility in individuals with substantial SIBs, providing a solid basis to suggest a distinct underlying neural signature. This is not as clear in the case of other (non-SIB) RRBs, which show higher volatility in only a few (4/31) of the relevant FCs.

The prevalence of high volatility in SIBs for connections in the MF and HF bands may be related to imbalance in the excitation and inhibition of neural activity in ASD; excess of the former may cause more frequent brain activity and induce higher connection frequencies [32]. Coupled with the strong mapping to areas involved in emotions and rewards processing, this is in general agreement with [33], a non-ASD-specific study which established that connections within limbic networks distribute over a wide range (0.01–0.14 Hz), whereas those in cortical networks are mostly limited to lower frequencies (0.01–0.06 Hz). In our results, however, there is no indication of the positive correlation found in [33] between connectivity decay along frequency, and physical distance.

The cerebellum is heavily involved throughout all sub-categories of relevant FCs (Table II). It has important contributions to cognitive functions and the "social brain", and its involvement in ASD is well established – see the recent review in [34]. In [35], for example, abnormalities were found in the structural covariation of the cerebellum and the amygdala, as well as sensory-related cortical structures in ASD. The morphology of the cerebellar vermis has also been studied extensively in relation to ASD, albeit with mixed results [36]. It has been associated with social and affective processing, as well as with the modulation of emotion in both physiological and behavioral studies [36-37]. Also of interest, [38] reports connectivity alterations in ASD broadly affecting the cerebellum, visual, and sensory-motor networks, which may underlie impairments in multisensory integration.

The FC between right amygdala and cerebellum vermis 1/2 (AMYG.R–CBLv12, SS2 > CTL, HF) stands out, as the amygdala also has a well-established relevance in ASD. It is involved in the perception of emotions and the control of aggression, and plays a critical role in the context of the social difficulties that characterize ASD [39]. Studies have shown that ASD groups have decreased connectivity between regions of the "social brain" [40], or no activity in the amygdala when making mentalistic inferences from the eyes [41]. The work in [42] concluded that activity in the amygdala comprises a combination of emotional and social signals: across an ASD group anxiety levels and core ASD deficits were, respectively, positively and negatively correlated with amygdala activity. In contrast, [43] found increased activity in the amygdala of adults with ASD, which may contribute to the social deficits which are characteristic of the disorder.

In our context, the higher volatility associated with SIBs for the FC between amygdala and cerebellum vermis may suggest mechanisms to alleviate emotional stress via frequently-varying engagement of the cerebellum in motor control.

Table II – Summary of relevant FCs, at ≥ 40% incidence under resampling scheme. Entries for 'X vs Y' columns are Cohen's-*d* effect size. Positive values imply Volatility(X) > Volatility(Y). Frequency bands: H (high), M (medium), L (low). SS2: ASD subjects at or above threshold for RBS-R subscale SS2; SS0: ASD subjects below threshold for SS2, but at or above threshold for any of SS1, SS3, SS4, SS5 or SS6; CTL: controls.

| FC Designation, per AAL Atlas [28] | | | Freq | SS2 vs CTL | SS2 vs SS0 | SS0 vs CTL |
|---|---|---|---|---|---|---|
| 5 | 102 | SFGorb.L   CBL7.R | H | 0.94 | | |
| 6 | 102 | SFGorb.R   CBL7.R | H | 0.91 | | |
| 10 | 102 | MFGorb.R   CBL7.R | M | 0.90 | | |
| 15 | 85 | IFGorb.L   MTG.L | M | 0.93 | | |
| 15 | 102 | IFGorb.L   CBL7.R | M | 0.85 | | |
| 21 | 38 | OLF.L   HIP.R | H | 0.95 | | |
| 23 | 95 | SFGmed.L   CBL3.L | M | 1.07 | | |
| 25 | 87 | MedFGorb.L   TPOmed.L | M | 0.99 | | |
| 26 | 87 | MedFGorb.R   TPOmed.L | M | 1.10 | | |
| 28 | 87 | REC.R   TPOmed.L | M | 1.01 | | |
| 35 | 87 | PCG.L   TPOmed.L | M | 0.88 | | |
| 38 | 109 | HIP.R   CBLv12 | H | 0.97 | | |
| 42 | 109 | AMYG.R   CBLv12 | H | 0.87 | | |
| 50 | 56 | SOG.R   FFG.R | L | 0.92 | | |
| 65 | 95 | ANG.L   CBL3.L | M | 0.91 | | |
| 71 | 100 | CAU.L   CBL6.R | H | 0.96 | | |
| 85 | 87 | MTG.L   TPOmed.L | M | 0.88 | | |
| 86 | 87 | MTG.R   TPOmed.L | M | 0.97 | | |
| 9 | 102 | MFGorb.L   CBL7.R | H | | 1.02 | |
| 10 | 18 | MFGorb.R   ROL.R | M | | 1.14 | |
| 15 | 90 | IFGorb.L   ITG.R | H | | 1.05 | |
| 15 | 91 | IFGorb.L   CBLc1.L | M | | 0.98 | |
| 15 | 92 | IFGorb.L   CBLc1.R | M | | 1.11 | |
| 102 | 107 | CBL7.R   CBL10.L | M | | 0.98 | |
| 10 | 101 | MFGorb.R   CBL7.L | M | 1.13 | 0.94 | |
| 15 | 94 | IFGorb.L   CBLc2.R | H | 0.85 | 1.04 | |
| 24 | 95 | SFGmed.R   CBL3.L | M | 1.16 | 1.15 | |
| 64 | 101 | SMG.R   CBL7.L | H | | –0.95 | |
| 88 | 105 | TPOmed.R   CBL9.L | M | | –1.01 | |
| 105 | 114 | CBL9.L   CBLv8 | H | | –1.07 | |
| 96 | 115 | CBL3.R   CBLv9 | H | | | 0.83 |

| FCs per Freq. Band: | Avg. Dist. [mm] | % Contralat. | FC totals, per type: | Avg. Dist. [mm]: | % Contralat.: |
|---|---|---|---|---|---|
| H: 12 (38.7%) | 83.87 | 58.3 | SS2 > CTL: 18 | 80.22 | 38.9 |
| M: 18 (58.1%) | 88.31 | 50.0 | SS2 > SS0: 6 | 99.86 | 66.7 |
| L: 1 (3.2%) | 66.44 | 0.0 | (SS2 > CTL) & (SS2 > SS0): 3 | 120.24 | 100.0 |
| | | | SS2 < SS0: 3 | 76.80 | 66.7 |
| | | | SS0 > CTL: 1 | 28.07 | 0.0 |

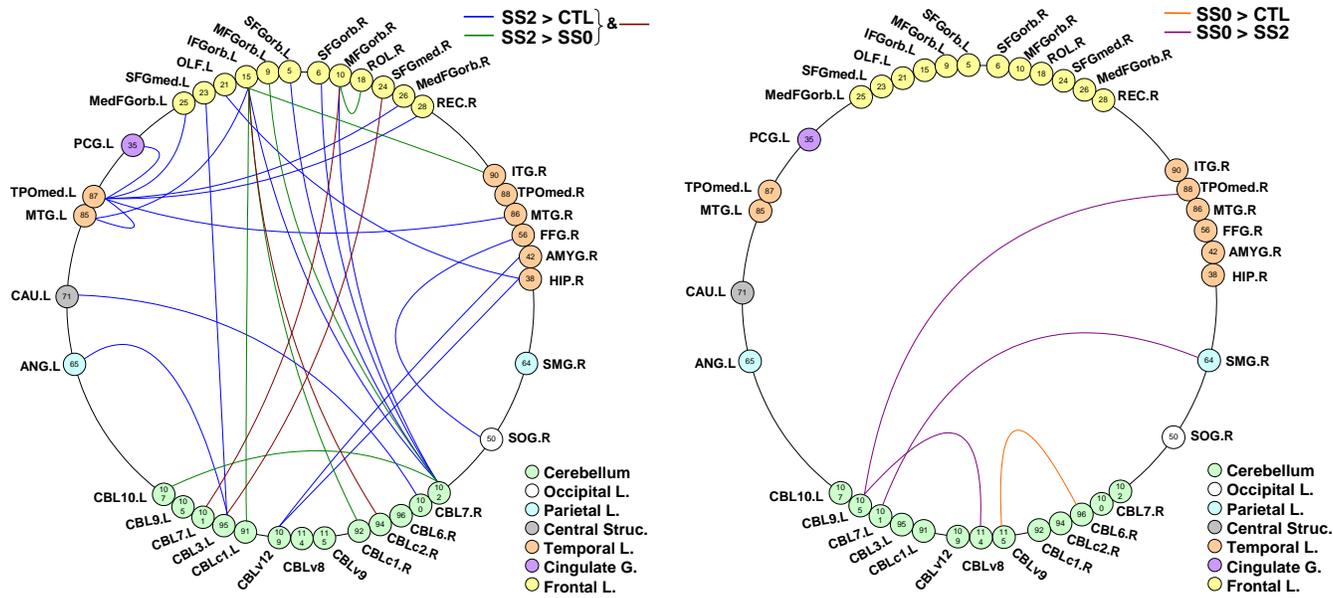

Fig. 2 – Mapping of all relevant FCs broken by comparison pairs. SS2: ASD subjects at or above threshold for RBS-R's SS2 subscale; SS0: ASD subjects below threshold for SS2, but at or above threshold for any of SS1, SS3, SS4, SS5 or SS6; CTL: controls.

Table III – Mapping of ROIs in relevant FCs to brain networks. All entries in [%]. HF, MF, LF: frequency bands. AAL: Atlas reference. DMN: default-mode; VIS: visual; SEN: sensorimotor; CBL: cerebellum; SAL: salience; OTH: others.

|         | DMN  | VIS   | SEN | CBL  | SAL  | OTH  |
|---------|------|-------|-----|------|------|------|
| HF      | 33.3 | 2.1   | 0.0 | 50.0 | 10.4 | 4.2  |
| MF      | 54.2 | 4.2   | 0.0 | 30.6 | 0.0  | 11.1 |
| LF      | 0.0  | 100.0 | 0.0 | 0.0  | 0.0  | 0.0  |
| **Overall** | **44.4** | **6.5** | **0.0** | **37.1** | **4.0** | **8.1** |
| AAL     | 28.4 | 13.8  | 9.5 | 22.4 | 8.6  | 17.2 |

The left inferior frontal gyrus (orbital part), IFGorb.L, which plays a role in all sub-categories with higher volatility in SS2, merits some attention. Associations between the orbitofrontal cortex and repetitive and stereotyped behaviors were reported in [44], while [45] detected associations between the orbitofrontal cortex and the visual system in ASD, and suggested that structural abnormalities in the former could affect the functioning of the amygdala. Structural abnormalities in the orbitofrontal cortex have also been reported in [46-47].

In the sub-category SS2 > CTL the left medial temporal gyrus of the temporal pole (TPOmed.L) clearly stands out, including connections to right gyrus rectus (REC.R) and bilaterally to MedFGorb in the frontal lobes, and bilaterally to MTG in the temporal lobes. Notably, abnormalities in the temporal poles of ASD subjects with comorbid developmental delay have been reported [48]: larger gray matter volumes, correlated with severity of impairments. The caudate nucleus, gyrus rectus (part of the medial prefrontal cortex) and hippocampus are also interesting on account of their involvement in the reward circuit, clearly important in our context as general reward dysfunction in ASD has been documented extensively [49], and in particular with regard to SIBs [50-51]. On balance, abnormalities in the volatility of FCs involving the reward circuit seem to contribute to a neural signature underlying self-injury in ASD.

The sub-category SS2 > SS0 should in principle provide a different angle to differentiate underlying signatures for SIBs and other RRBs. However, most of the ROIs involved (multiple areas of the cerebellum, MFGorb and IFGorb) are also involved in SS2 > CTL, making any such distinction difficult. We do note the two sub-categories differ in that connection distances, on average, are longer for SS2 > SS0 (99.86 vs 80.22 mm, Table II), with a higher proportion being contralateral (66.7% vs 38.9%). There is no clear association between this and the frequency bands over which the FCs occur: for both sub-categories two-thirds of them are in MF.

While the small sub-category of FCs yielding higher volatility for SS2 relative to both CTL and SS0 offers little insight, it is interesting in that all FCs involve the cerebellum and parts of the frontal gyrus, are contralateral, and have the longest average distance between connections. On the other hand, a potential marker for non-SIB RRBs may be found in higher volatility in FCs involving CBL9.L, CBLv8, CBL3.R, CBLv9, and SMG.R – SS0 > SS2, and SS0 > CTL (Table II).

There is a sizable (~20%) overlap of ROIs in relevant FCs with those from [26]. Additionally, the FC between right-amygdala and cerebellum vermis 1/2 (AMYG.R–CBLv12, SS2 > CTL, HF, Table II), has a closely related counterpart in the FC between left-amygdala and the same region of the cerebellum (AMYG.L–CBLv12). Per [26], the latter was a particularly strong finding of stronger connectivity in severe cases of RRBs compared to controls, detected by both static and windowed correlations, and preprocessing with and without global signal regression. Any degree of concordance between the studies is interesting on account of (a) their being based on different data sets (ABIDE-II vs ABIDE-I), (b) different preprocessing, and (c) different metrics to assess RRBs (RBS-R vs ADI-R-RRB).

Some limitations must be noted. Despite the seemingly large pool of potential subjects offered by the multisite database used, once all phenotypical restrictions are accounted for the resulting sample size is relatively small. Additionally, the range of RBS-R scores for all subscales is heavily biased toward the lower (milder) end, further restricting sample sizes if subscale groups limited to very-high severity levels were to be formed. The resulting low sensitivity of any statistical analysis conducted on such groups would prevent meaningful insight into potentially different underlying neural signatures as a function of severity level. This emphasizes the gross under-representation of severe cases of RRBs in ASD studies.

On balance, the volatility measure proposed to analyze the dynamics of resting-state functional connectivity in time-frequency space appears promising as a tool to study RRBs in ASD. This work contributes to the identification of neural substrates potentially underlying such behaviors, and may help target interventions. Overall findings indicate a preponderance of relevant FCs between areas associated with rewards and emotions processing, in medium- and high-frequency bands, for the most part yielding higher volatility in subjects with substantial SIBs.

ACKNOWLEDGMENT

The author received no funding for this work, which was carried out in a private capacity, and declares no competing interests. All scan data used are from a public repository. Output data sets are available at [52].

APPENDIX A – Complete summary of all relevant FCs, for analyses with full-size groups ($N_{SS0}$=29, $N_{SS2}$=18, $N_{CTL}$=50). Frequency bands: HF (high), MF (medium), LF (low). m1 and m2: mean measures for A and B, respectively, in 'A vs B' comparison pair. *d*: Cohen's-*d* effect size; Pwr: t-test's power (sensitivity). RS: re-sampling (100 runs, 2 randomly-selected subjects excluded from each of SS0, SS2 and CTL for each run), % incidence for current approach (≥ 40% in red). RS-REF: % incidence for preceding approach [25], for relevant FCs in common with current approach.

| One-way ANOVA, two-sample t-test: | | | | | | | | | | | | Linear reg. model: Grps (G), Sites (S), Age (A) | | | | | | RS | RS-REF |
|---|---|---|---|---|---|---|---|---|---|---|---|---|---|---|---|---|---|---|---|
| ROI-1 | ROI-2 | | | Frq | p | F-stat | Conf. Interval | | m1-m2 | d | PWR | GF | Gp | SF | Sp | AF | Ap | % | % |
| **SS0 vs SS2** | | | | | | | | | | | | | | | | | | | |
| 9 | 102 | MFGorb.L | CBL7.R | HF | 0.0043 | 5.77 | -4.38e-01 | -3.85e-02 | -2.38e-01 | -1.02 | 0.80 | 5.59 | 0.0052 | 2.12 | 0.1036 | 2.26 | 0.1359 | **60** | |
| 15 | 90 | IFGorb.L | ITG.R | HF | 0.0029 | 6.23 | -4.76e-01 | -4.86e-02 | -2.62e-01 | -1.05 | 0.83 | 6.01 | 0.0035 | 0.24 | 0.8660 | 0.02 | 0.8837 | **77** | |
| 15 | 94 | IFGorb.L | CBLc2.R | HF | 0.0022 | 6.54 | -5.17e-01 | -5.06e-02 | -2.84e-01 | -1.04 | 0.82 | 6.04 | 0.0035 | 0.89 | 0.4517 | 0.85 | 0.3589 | **73** | 26 |
| 64 | 101 | SMG.R | CBL7.L | HF | 0.0064 | 5.33 | 2.25e-02 | 4.12e-01 | 2.17e-01 | 0.95 | 0.74 | 5.10 | 0.0080 | 1.27 | 0.2890 | 0.00 | 0.9647 | **49** | |
| 105 | 114 | CBL9.L | CBLv8 | HF | 0.0021 | 6.57 | 4.86e-02 | 4.35e-01 | 2.42e-01 | 1.07 | 0.84 | 7.43 | 0.0010 | 0.83 | 0.4831 | 0.52 | 0.4708 | **87** | 51 |
| 10 | 18 | MFGorb.R | ROL.R | MF | 0.0012 | 7.24 | -7.00e-01 | -1.01e-01 | -4.00e-01 | -1.14 | 0.89 | 6.04 | 0.0035 | 1.27 | 0.2878 | 0.95 | 0.3312 | **82** | 92 |
| 10 | 101 | MFGorb.R | CBL7.L | MF | 0.0004 | 8.60 | -6.86e-01 | -3.43e-02 | -3.60e-01 | -0.94 | 0.73 | 10.67 | 0.0001 | 1.17 | 0.3257 | 2.01 | 0.1599 | **45** | 71 |
| 15 | 91 | IFGorb.L | CBLc1.L | MF | 0.0040 | 5.87 | -8.43e-01 | -5.98e-02 | -4.51e-01 | -0.98 | 0.77 | 5.47 | 0.0057 | 0.41 | 0.7496 | 0.15 | 0.6984 | **58** | |
| 15 | 92 | IFGorb.L | CBLc1.R | MF | 0.0015 | 7.01 | -9.19e-01 | -1.21e-01 | -5.20e-01 | -1.11 | 0.88 | 6.94 | 0.0016 | 0.23 | 0.8781 | 0.60 | 0.4413 | **92** | 43 |
| 24 | 95 | SFGmed.R | CBL3.L | MF | 0.0001 | 9.81 | -6.34e-01 | -9.36e-02 | -3.64e-01 | -1.15 | 0.90 | 7.50 | 0.0010 | 2.63 | 0.0552 | 0.31 | 0.5821 | **69** | |
| 28 | 87 | REC.R | TPOmed.L | MF | 0.0013 | 7.17 | -5.52e-01 | -2.50e-01 | -2.89e-01 | -0.93 | 0.72 | 10.36 | 0.0001 | 2.24 | 0.0890 | 0.20 | 0.6581 | 33 | 47 |
| 65 | 90 | ANG.L | ITG.R | MF | 0.0090 | 4.95 | -6.88e-01 | -3.02e-02 | -3.59e-01 | -0.93 | 0.72 | 4.74 | 0.0110 | 2.13 | 0.1014 | 0.07 | 0.7924 | 29 | |
| 88 | 105 | TPOmed.R | CBL9.L | MF | 0.0032 | 6.10 | 6.54e-02 | 7.85e-01 | 4.25e-01 | 1.01 | 0.79 | 6.57 | 0.0022 | 0.57 | 0.6371 | 1.33 | 0.2520 | **66** | |
| 102 | 107 | CBL7.R | CBL10.L | MF | 0.0035 | 6.02 | -6.79e-01 | -4.54e-02 | -3.62e-01 | -0.98 | 0.76 | 6.71 | 0.0019 | 0.92 | 0.4358 | 0.08 | 0.7764 | **49** | |
| **SS0 vs CTL** | | | | | | | | | | | | | | | | | | | |
| 96 | 115 | CBL3.R | CBLv9 | HF | 0.0013 | 7.18 | 3.27e-02 | 2.99e-01 | 1.66e-01 | 0.83 | 0.85 | 6.29 | 0.0028 | 1.15 | 0.3333 | 0.81 | 0.3708 | **60** | |
| **SS2 vs CTL** | | | | | | | | | | | | | | | | | | | |
| 5 | 102 | SFGorb.L | CBL7.R | HF | 0.0027 | 6.30 | 4.01e-02 | 4.47e-01 | 2.44e-01 | 0.94 | 0.81 | 5.08 | 0.0082 | 1.52 | 0.2145 | 0.19 | 0.6655 | **58** | |
| 6 | 102 | SFGorb.R | CBL7.R | HF | 0.0048 | 5.64 | 3.30e-02 | 4.33e-01 | 2.33e-01 | 0.91 | 0.79 | 4.69 | 0.0115 | 0.86 | 0.4651 | 0.60 | 0.4397 | **51** | |
| 15 | 94 | IFGorb.L | CBLc2.R | HF | 0.0022 | 6.54 | 1.96e-02 | 4.47e-01 | 2.33e-01 | 0.85 | 0.73 | 6.04 | 0.0035 | 0.89 | 0.4517 | 0.85 | 0.3589 | **44** | 31 |
| 21 | 38 | OLF.L | HIP.R | HF | 0.0021 | 6.60 | 4.24e-02 | 4.44e-01 | 2.43e-01 | 0.95 | 0.82 | 4.89 | 0.0096 | 2.17 | 0.0976 | 0.01 | 0.9122 | **52** | 68 |
| 38 | 109 | HIP.R | CBLv12 | HF | 0.0021 | 6.57 | 3.87e-02 | 3.67e-01 | 2.03e-01 | 0.97 | 0.84 | 5.33 | 0.0065 | 0.29 | 0.8295 | 0.10 | 0.7547 | **70** | |
| 42 | 109 | AMYG.R | CBLv12 | HF | 0.0079 | 5.10 | 2.00e-02 | 3.63e-01 | 1.91e-01 | 0.87 | 0.75 | 4.63 | 0.0122 | 0.93 | 0.4277 | 0.35 | 0.5575 | **41** | 63 |
| 71 | 100 | CAU.L | CBL6.R | HF | 0.0027 | 6.31 | 4.84e-02 | 4.62e-01 | 2.55e-01 | 0.96 | 0.84 | 4.86 | 0.0099 | 2.30 | 0.0829 | 0.17 | 0.6802 | **49** | |
| 10 | 101 | MFGorb.R | CBL7.L | MF | 0.0004 | 8.60 | 1.34e-01 | 7.30e-01 | 4.32e-01 | 1.13 | 0.94 | 10.67 | 0.0001 | 1.17 | 0.3257 | 2.01 | 0.1599 | **100** | 100 |
| 10 | 102 | MFGorb.R | CBL7.R | MF | 0.0057 | 5.46 | 4.49e-02 | 6.46e-01 | 3.45e-01 | 0.90 | 0.78 | 6.98 | 0.0015 | 1.09 | 0.3581 | 0.17 | 0.6828 | **68** | 94 |
| 15 | 85 | IFGorb.L | MTG.L | MF | 0.0043 | 5.76 | 5.96e-02 | 6.76e-01 | 3.68e-01 | 0.93 | 0.81 | 5.01 | 0.0086 | 0.18 | 0.9130 | 1.03 | 0.3122 | **52** | |
| 15 | 102 | IFGorb.L | CBL7.R | MF | 0.0073 | 5.19 | 3.41e-02 | 7.81e-01 | 4.07e-01 | 0.85 | 0.73 | 5.09 | 0.0080 | 0.45 | 0.7183 | 0.65 | 0.4216 | **44** | |
| 23 | 95 | SFGmed.L | CBL3.L | MF | 0.0008 | 7.72 | 9.87e-02 | 6.29e-01 | 3.64e-01 | 1.07 | 0.92 | 7.31 | 0.0011 | 2.48 | 0.0665 | 0.04 | 0.8326 | **83** | |
| 24 | 95 | SFGmed.R | CBL3.L | MF | 0.0001 | 9.81 | 1.20e-01 | 6.15e-01 | 3.68e-01 | 1.16 | 0.96 | 7.50 | 0.0010 | 2.63 | 0.0552 | 0.31 | 0.5821 | **69** | |
| 25 | 87 | MedFGorb.L | TPOmed.L | MF | 0.0019 | 6.70 | 7.59e-02 | 6.42e-01 | 3.59e-01 | 0.99 | 0.86 | 8.41 | 0.0004 | 1.77 | 0.1591 | 0.31 | 0.5769 | **84** | 75 |
| 26 | 87 | MedFGorb.R | TPOmed.L | MF | 0.0006 | 8.03 | 1.31e-01 | 7.76e-01 | 4.53e-01 | 1.10 | 0.93 | 8.50 | 0.0004 | 0.70 | 0.5552 | 0.01 | 0.9128 | **97** | 93 |
| 27 | 71 | REC.L | CAU.L | MF | 0.0074 | 5.17 | 2.98e-02 | 6.29e-01 | 3.29e-01 | 0.86 | 0.73 | 5.45 | 0.0058 | 1.72 | 0.1691 | 0.04 | 0.8443 | 39 | |
| 28 | 87 | REC.R | TPOmed.L | MF | 0.0013 | 7.17 | 7.20e-02 | 5.55e-01 | 3.13e-01 | 1.01 | 0.88 | 10.36 | 0.0001 | 2.24 | 0.0890 | 0.20 | 0.6581 | **80** | 67 |
| 35 | 87 | PCG.L | TPOmed.L | MF | 0.0076 | 5.14 | 3.34e-02 | 5.59e-01 | 2.96e-01 | 0.88 | 0.76 | 5.12 | 0.0078 | 0.09 | 0.9653 | 1.89 | 0.1730 | **46** | |
| 58 | 96 | PostCG.L | CBL3.R | MF | 0.0080 | 5.09 | 2.08e-02 | 5.53e-01 | 2.87e-01 | 0.84 | 0.71 | 4.54 | 0.0132 | 0.14 | 0.9361 | 3.80 | 0.0542 | 21 | |
| 65 | 95 | ANG.L | CBL3.L | MF | 0.0054 | 5.51 | 4.60e-02 | 6.01e-01 | 3.23e-01 | 0.91 | 0.79 | 4.84 | 0.0101 | 1.76 | 0.1606 | 0.04 | 0.8514 | **56** | 47 |
| 85 | 87 | MTG.L | TPOmed.L | MF | 0.0055 | 5.50 | 3.94e-02 | 6.62e-01 | 3.51e-01 | 0.88 | 0.76 | 5.83 | 0.0041 | 1.57 | 0.2025 | 0.54 | 0.4649 | **49** | |
| 86 | 87 | MTG.R | TPOmed.L | MF | 0.0028 | 6.28 | 7.62e-02 | 6.96e-01 | 3.86e-01 | 0.97 | 0.85 | 5.60 | 0.0051 | 0.28 | 0.8392 | 0.05 | 0.8248 | **70** | 63 |
| 102 | 107 | CBL7.R | CBL10.L | MF | 0.0035 | 6.02 | 2.50e-02 | 6.05e-01 | 3.15e-01 | 0.85 | 0.72 | 6.71 | 0.0019 | 0.92 | 0.4358 | 0.08 | 0.7764 | 35 | |
| 50 | 56 | SOG.R | FFG.R | LF | 0.0039 | 5.90 | 5.31e-02 | 6.39e-01 | 3.46e-01 | 0.92 | 0.80 | 4.73 | 0.0111 | 2.16 | 0.0983 | 1.48 | 0.2268 | **49** | |